\documentstyle[12pt,draft]{article}

 \newcommand {\bi} {\bibitem}
 \newcommand {\be} {\begin{equation}}
 \newcommand {\bd} {\begin{math}}
\newcommand {\bea} {\begin{eqnarray} \nonumber }
\newcommand {\ee} {\end{equation}}
\newcommand {\ed} {\end{math} }
\newcommand {\eea} {\end{eqnarray}}
 
 \newcommand {\si} {\sigma}
\newcommand {\de} {\delta}
\newcommand {\De} {\Delta}
\newcommand {\ga} {\gamma}
\newcommand {\la} {\lambda}

 \newcommand {\al} {\alpha}

\newcommand {\ba} {\overline}

\def \form#1 {eq. (\ref{#1}) }
\def \parziale#1#2  {{\partial {#1} \over \partial {#2}}}
 
\begin{document}

\title{On the mean field approach to glassy systems}
\author{  Giorgio Parisi\\
Dipartimento di Fisica, Universit\`a {\em La  Sapienza},\\ 
INFN Sezione di Roma I \\
Piazzale Aldo Moro, Rome 00185}
\maketitle

\begin{abstract}
In these lectures I will study some properties that are shared by many glassy systems.  I will shown 
how some of these properties can be understood in the framework of the mean field approach based on 
the replica method and I will discuss which are the difficulties which we have to face when we apply 
the replica method to realistic short range models.  We finally present a partially successfully 
application of the replica method to soft sphere glasses.
\end{abstract}

\section {Introduction}

In these lecture I will study some systems which present some of the relevant characteristics of 
glasses.  The aim is to construct simple microscopic models which can be studied in details and 
still behave in an interesting way.  I will start from the simplest models, where only some of the 
observed characteristics are reproduced and we will later go to more complex systems.

The basic idea is to learn as much it is possible from mean field theory results and to try to 
understand the extent of the overlap among the rather complex phenomenology displayed by glasses and 
the exact results which are obtained for model with infinite range interaction.  It is remarkable 
that the two cases have many points in common (with a few very interesting differences!) so that it 
seems to be a quite natural and reachable goal to construct a theory of glasses based on  mean 
field tools.  This construction is in progress and here I will review our present understanding.

This note is organized as follows: after this introduction we recall some of the main results which 
have been obtained in the framework of the mean field theory, both for systems with and {\it 
without} quenched disorder.  In section III I will show which are the difficulties to extend these 
results to short range models, which properties are maintained and which are modified.  Finally in 
the last section I will stress some of the properties of real glasses and I will describe a first 
tentative of doing esplicite computations for a soft sphere glass.

\section{Mean field results}
In these recent years there have been many progresses on the understanding of the behaviour of 
glassy systems in the mean field approximation.  The mean field approximation is correct when the 
range of the interaction is infinite and this property allows us to write self consistent equations 
whose solution gives the solution of the model.  The techniques that can be used are of various 
types, the replica method, a direct probabilistic method based of the cavity (Bethe) equations
\cite{mpv,parisibook2} and the direct study of the dynamical equations \cite {CUKU,CRISO}.

 The main results have been the following:
\begin{itemize}
\item Model with random quenched disorder have been well understood both from 
the equilibrium and from the dynamical point of view.

\item Some of the results obtained for systems with random quenched disorder 
have been extended to non disordered system.  This step is crucial in order to 
have the possibility of extending these models to real glasses. 
\end{itemize}
Let us see in details the results that have been obtained.

\subsection{Disordered systems}

In this case the thermodynamical properties at equilibrium can be computed using the replica method 
\cite{mpv,parisibook2}.  A typical example of a model which can be solved with the replica method is 
a spin model with $p$ spin interaction.  The Hamiltonian we consider depends on some control 
variables $J$, which have a Gaussian distribution and play the same role of the random energies of 
the REM (i.e.  the random energy model \cite{REM,ISOPI}) and by the spin variable $\si$.  For 
$p=1,2,3$ the Hamiltonian is respectively
\begin{eqnarray}
H^1_J(\si)= \sum_{i=1,N} J_i \si_i\\
H^2_J(\si)= \sum_{i,k=1,N}' J_{i,k} \si_i \si_k\\
H^3_J(\si)= \sum_{i,k,l=1,N}' J_{i,k,l} \si_i \si_k \si_l \nonumber
\end{eqnarray}
where the primed sum indicates that all the indices are different.  The 
$N$ variables $J$ must have a variance of $O(N^{(1-p)/2})$ if we want to have a non trivial 
thermodynamical limit.  The variables $\si$ are usual Ising spins, which take the values $\pm 1$.  
From now on we will consider only the case $p>2$.

In the replica approaches one assumes that at low temperatures the phase space 
breaks into many valleys, (i.e.  regions separated by high barriers in free 
energy).  One also introduces the overlap among valley as
\be
q(\al,\ga)\equiv {\sum_{i=1,N} \si^\al_i \si^\ga_i \over  N},
\ee
where $\si^\al$ and $\si^\ga$ are two generic configurations in the valley $\al$ and $\ga$ 
respectively.

In the simplest version of this method \cite{GROMEZ,GARDNER} one introduces the typical overlap of 
two configurations inside the same valley (sometimes denoted by $q_{EA}$.  Something must be said 
about the distribution of the valleys.  Only those which have minimum free energy are relevant for 
the thermodynamics.  One finds that these valleys have zero overlap and have the following 
probability distribution of {\em total} free energy of each valley:
\be P(F)\propto \exp(\beta m (F-F_0)),  \label{FREE} \ee
where $F_{0}$ is the total free energy of the valley having lower free energy.

Indeed the average value of the free energy can be written in a self consistent way as function of 
$m$ and $q$ ($f(q,m)$) and the value of these two parameters can be found as the solution of the 
stationarity equations:
\be \parziale{f}{m} =\parziale{f}{q} =0.  \ee

The quantity $q$ is of order $1-\exp(-A\beta p)$ for large $p$, while the parameter $m$ is 1 at the 
critical temperature, and has a nearly linear behaviour al low temperature.  The only difference is 
that $m$ is no more strictly linear as function of the temperature.

The thermodynamical properties of the model are the same as is the Random Energy Model (indeed we 
recover the REM when $p\to\infty$): there is a transition at $T_{c}$ with a discontinuity in the 
specific heat, with no divergent susceptibilities.

A very interesting finding is that if we consider the infinite model and we cool it starting at high 
temperature, there is a transition at a temperature $T_{D}>T_{c}$ \cite{CUKU,CRISO}.  At temperatures 
less than $T_{D}$ the system is trapped in a metastable state.  The correlation time (not the 
equilibrium susceptibilities) diverges at $T_{D}$ and the mode-mode coupling become exact in this 
region \cite{BCKM} .

It was suggested some time ago \cite{KT1,KT2,KT3} that these properties of the $p$-spin model 
strongly hint that this model may be considered a mean field realization of a glassy system and it 
should share with real glasses the physical origine of this behaviour.

It is interesting to note that although in the original approach the dynamical transition was found 
by using explicitly the equation of motions now we have techniques which allow the determination of 
the dynamical transition and of some of the properties of the exponentially large times needed to 
escape from a metastable state using only equilibrium computations \cite{KLEIN,FP,MONA,CAGIAPA}.  
The main technical tool consists in computing the properties of a system with coupled replicas
\cite{KPVI}. 

\subsection{Model without quenched disorder}
We could ask how much of the previous results can be carried to models without quenched results.  It 
has been found in the framework of the mean field theory (i.e.  when the range of the interaction is 
infinite), that there a the partial equivalence of Hamiltonians with quenched and random disorder.  
More precisely it often possible to find Hamiltonians which have the same properties (at least in a 
region of the phase space) of the Hamiltonian without disorder \cite{MPR} - \cite{CKPR} .  An 
example of this intriguing phenomenon is the following.

The configuration space of our model is given by $N$ Ising spin variables \cite{MPR}.  We consider 
the following Hamiltonian
\begin{eqnarray}
H=\sum_{i=1,N}|B_i|^2-1|^2, \\
\mbox{where} \ \ \ B_i=\sum_{k=1,N} R_{i,k} \si_k.
\end{eqnarray}
Here $R$ is an unitary matrix, i.e.
\be
\sum_{k=1,N}R_{i,k}\ba{R_{k,m}}=\de_{i,m}.
\ee

We could consider two different cases \cite{MPR} :
\begin{itemize}
\item The matrix $R$ is a random orthogonal matrix.
\item The matrix $R$ is given by
\be
R(k,m) ={\exp (2 \pi i \ k m) \over N^{1/2}}
\ee
In other words $B$ is the Fourier transform of $\si$.
\end{itemize}

The second case is a particular instance of the first one, exactly in the same way that a sequence 
of all zeros is a particular instance of a random sequence.

The first model can be studied using the replica method and one finds results 
very similar to those of the $p$-spin model we have already studied.

Now it can be proven that the statistical properties of the second model are identical to those of 
the first model, with however an extra phase.  In the second model (at least for some peculiar value 
of $N$, e.g.  $N$ prime \cite{MIGLIO,MPR,BGU} ) there are configurations which have exactly zero 
energy.  These configuration form isolated valleys which are separated from the others, but have 
much smaller energy and they have a very regular structure (like a crystal).  An example of these 
configurations is
\be \si_k \equiv_{ \bmod N }k^{(N-1)/2} \ee 
(The property $k^{(N-1)}\equiv 1$ for prime \bd N \ed , implies that in the previous equations 
$\si_k=\pm 1$).  Although the sequence $\si_k$ given by the previous equation is apparently random, 
it satisfies so many identities that it must be considered as an extremely ordered sequence (like a 
crystal).  One finds out that from the thermodynamical point of view it is convenient to the system 
to jump to one of these ordered configurations at low temperature.  More precisely there is a first 
order transition (like a real crystalization transition) at a temperature, which is higher that the 
dynamical one.

If the crystallisation transition is avoided by one of the usual methods, (i.e.  explicit 
interdiction of this region of phase space or sufficient slow cooling), the properties of the second 
model are exactly the same of those of the first model.  Similar considerations are also valid for 
other spin models
\cite{Frhe,MPRII,MPRIII} or for models of interacting particles in very large dimensions, where the
effective range of the force goes to infinity \cite{CKPR,CKMP,PA97,PR} .

We have seen that when we remove the quenched disorder in the Hamiltonian we find  a quite positive 
effect: a crystallisation transition appears like in some real systems.  If we neglect 
crystalization, which is absent for some values of $N$, no new feature is present in system without 
quenched disorder.  

These results are obtained for long range systems. As we shall see later the equivalence of short 
range systems with and without quenched disorder is an interesting and quite open problem.

\section{Short range models}

The interest of the previous argument would be to much higher if we could apply them to short range 
models.  In order to discuss this point it is convenient to consider two classes of models:
\bea
H=\sum_{x,y,z}J(x,y,z)\si(x)\si(y)\si(z)\\
H=\sum_{x,y,z}J(x-y,x-z)\si(x)\si(y)\si(z).
\eea

In both cases the sum is restricted to triplets spins which are at distance less than $R$ (i.e.  
$|x-y|<R$, $|x-z|<R$, $|z-y|<R$).  In the first case the model is not translational invariant while 
in the second case the Hamiltonian is translational invariant.  Moreover in the second case the 
Hamiltonian depends only on a {\em finite} number of $J$ also when the volume goes to infinity, so 
that the free energy density will fluctuates with $J$.  The second model looks much more similar to 
real glasses than the first one.  A careful study of the difference among these two models has not 
yet been done.  In any case it is clear that the phenomenon of metastability which we have seen in 
the previous section for the infinite range models is definitely not present in these short range 
models.  No metastable states with infinite mean life do exist in nature.

Let us consider a quite simple argument.  Let us suppose that the system may stay in phase (or 
valleys) which we denote as $A$ and $B$.  If the free energy density of $B$ is higher than that of 
$A$, the system can go from $B$ to $A$ in a progressive way, by forming a bubble of radius $R$ of 
phase $A$ inside phase $B$.  If the surface tension among phase $A$ and $B$ is finite, has happens 
in any short range model, for large $R$ the volume term will dominate the free energy difference 
among the pure phase $B$ and phase $B$ with a bubble of $A$ of radius $R$.  This difference is thus 
negative at large $R$, it maximum will thus be finite.  

In the nutshell a finite amount of free energy in needed in order to form a seed of phase $A$ 
starting from which the spontaneous formation of phase $A$ will start.  For example, if we take a 
mixture of $H_2$ and $O_2$ at room temperature, the probability of a spontaneous temperature 
fluctuation in a small region of the sample, which lead to later ignition and eventually to the 
explosion of the whole sample, is greater than zero (albeit quite a small number), and obviously it 
does not go to zero when the volume goes to infinity.

We have two possibilities open in positioning the mean field theory predictions of existence of real 
metastable states:
\begin{itemize}
  \item We consider the presence of these metastable state with {\sl infinite} mean life an artefact 
  of the mean field approximation and we do not pay attention to them.

  \item We notice that in the real systems there are metastable states with very large (e.g.  much 
  greater than one year) mean life.  We consider the {\sl infinite} time metastable states of the 
  mean field approximation as precursors of these {\sl finite} time metastable states.  We hope 
  (with reasons) that the corrections to the mean field approximation will give a finite (but large) 
  mean life to these states (how this can happen will be discussed later on).
\end{itemize}

Here we suppose that the second possibility is the most interesting and we proceed with the study of 
the system in the mean field approximation.  We have already seen that in a short range model we 
cannot have real metastable states.  Let us see in more details what happens.

Let us assume that $\De f<0$ is the difference in free energy among the 
metastable state and the stable state. 
Now let us consider a bubble of radius $R$ of stable state inside the metastable 
one. The free energy difference of such a bubble will be
\be
F(R) =-\De f \ V(R) - I(R)
\ee
where the interfacial free energy $I(R)$ can increase at worse as $\Sigma(R)$.  The quantities $V(R) 
\propto R^D$ and $\Sigma(R) \propto R^{D-1}$ are respectively the volume and the surface of the 
bubble; $\si$ is the surface tension, which can also be zero (when the surface tension is zero, we 
have $I(R)\propto R^\omega$, with $\omega<D-1$).
 
The value of $F(R)$ increases at small $R$, reaches a maximum a $R_c$, which in the case $\si\ne 0$ 
is of order $(\De f) ^{-1}$ and it becomes eventually negative at large $R$.  According to 
enucleation theory, the system goes from the metastable to the stable phase under the formation and 
the growth of such bubbles, and the time to form one of them is of order (neglecting prefactors)
\begin{eqnarray}
\ln (\tau) \propto \De f\  R^D \propto (\De f)^{(D-1)}\\
\tau \propto \exp( {A \over (\De f)^{(D-1)}})
\end{eqnarray}
where $A$ is constant dependent on the surface tension.
In the case of asymptotically zero surface tension we have
\begin{eqnarray}
\tau \propto \exp( {A \over \De f^\la})\\
\la={D \over \omega} -1
\end{eqnarray}

This argument for the non existence of metastable states can be naively applied here.  The 
metastable states of the the mean field approximation now do decay.  The dynamical transition 
becomes a smooth region which separates different regimes; an higher temperature regime where mode 
mean field predictions are approximately correct and a low temperature region where the dynamics is 
dominated by barriers crossing.  There is no region where the predictions of mean field theory are 
exact but mean field theory is only an approximated theory which describe the behaviour in a limited 
region of relaxation times (large, but not too large).

The only place where the correlation time may diverge is at that the thermodynamical transition 
$T_c$, whose existence seems to be a robust prediction of the mean field theory.  It follows that 
the only transition, both from the static and the dynamical point of view, is present at $T_c$.

In order to understand better what happens near $T_c$ we must proceed in a careful matter.  The 
nucleation phenomenon which is responsible of the decay of metastable states is of the same order of 
other non-perturbative corrections to the mean field behaviour, which cannot be seen in perturbation 
theory.  We must therefore compute in a systematic way all possible sources of non perturbative 
corrections.  This has not yet been done, but it should not be out of reach.

One of the first problem to investigate is the equivalence of systems with and without random 
disorder.  In systems with quenched disorder there are local inhomogeneities which correspond to 
local fluctuations of the critical temperature and may dominate the thermodynamics when we approach 
the critical temperature.

It is quite possible, that systems with and without quenched disorder, although they coincide in the 
mean field approximation, they will be quite different in finite dimension (e.g.  3) and have 
different critical exponents.  The scope of the universality classes would be one of the first 
property to assess.

It is not clear at the present moment if the strange one order and half transition is still present 
in short range model or if it is promoted to a bona fide second order transition.  If the transition 
remains of the order one and half (for example is conceivable that this happens only for systems 
without quenched disorder).  It could also possible that there is appropriate version of the 
enucleation theory which is valid near $T_c$ and predicts:

\be
\tau\propto \exp( {A \over (T-T_C)^\al}).
\ee

A first guess for $\al$ is $D-2$ \cite{KT3, KLEIN} , although other values, e.g.  2/3, are possible.  
A more detailed understanding of the static properties near $T_c$ is needed before we can do any 
reliable prediction.

\section{Toward realistic models}
\subsection{General considerations on glasses}

I would like now to shortly review the properties of real glasses, which, as we shall see, have many 
points in common with the systems that we have considered up to now.  As usual we must select which 
of the many characteristics of glasses we think are important and should be understood.  This is a 
matter of taste.  I believe that the the following facts are the main experimental findings about 
glasses that a successfully theory of glasses should explain:
\begin{enumerate}
\item If we cool the system below some temperature ($T_G$), its energy depends 
on the cooling rate in a significant way. We can visualize $T_G$ as the 
temperature at which the relaxation times become of the order of a hour.

\item No thermodynamic anomalies (i.e.  divergent specific heat or
susceptibilities) are observed: the entropy (extrapolated at ultraslow cooling) 
is a linear function of the temperature in the region where such an 
extrapolation is possible.  For finite value of the cooling rate the specific 
heat is nearly discontinuous.  Data are consistent with the possibility that the 
true equilibrium value of the specific heat is also discontinuous at a 
temperature $T_c$ lower than $T_G$.  This results comes mainly from systems 
which have also a crystal phase (which is reached by a different cooling 
schedule) under the very reasonable hypothesis that the entropy of the glass 
phase cannot be smaller than that of the crystal phase.

\item The relaxation time (and quantities related to it, e.g.  the 
viscosity, which by the Kubo formula is proportional to the integral of the correlation function of 
the stress energy tensor) diverges at low temperature.  In many glasses (the fragile ones) the 
experimental data can be fitted as
\begin{eqnarray} 
\tau =\tau_0 \exp(\beta B(T)),\\ 
B(T) \propto (T-T_c)^{-\la}, 
\end{eqnarray}
where $\tau_0 \approx 10^{-13} s$ is a typical microscopic time, $T_c$ is near to the value at which 
we could guess the presence of a discontinuity in the specific heat and the exponent $\la$ is of 
order 1.  The so called Vogel-Fulcher law \cite{VF} states that $\la=1$.  The precise value of $\la$ 
is not too well determine.  The value $1$ is well consistent with the experimental data, but 
different values are not excluded.
\end{enumerate}

Points 1 and 2 can be easily explained in the framework of the mean field approximation.  Point 3 is 
a new feature of short range models which is not present in the mean field approximation.  It is 
clearly connected to the non-existence of infinite mean life metastable states in a finite 
dimensional world.  One needs a more careful analysis in order to find out the origine of this 
peculiar behavior and obtain quantitative predictions.

It is interesting to find out if one can construct approximation directly for the glass transitions 
in liquids.  Some progresses have been done in the framework of mean field theory in the infinite 
dimensional cases.  Indeed the model for hard spheres moving on a sphere can be solved exactly in 
the high temperature phase when the dimension of the space goes to infinity in a suitable way
\cite {CKPR,CKMP,PR}.

\subsection{A first attempt to use the replica method for glasses}
One of the most interesting results is the suggestion that the replica method can be directly 
applied to real glasses.  The idea is quite simple \cite{MEPA}.  We assume that in the glassy phase a 
finite large system may state in different valleys (or states), labeled by $\al$.  The probability 
distribution of the free energy of the valley is given by eq.  (\ref{FREE}).  We can speak of a 
probability distribution because the shape of the valleys and their free energies depends on the 
total number of particles.  Each valley may be characterized by the density

\be
\rho(x)_{\al}\equiv< \rho(x)>_{\al}.
\ee
In this case we can define two correlation functions.
\bea
g(x)=\frac{\int dy < \rho(y) \rho (y+x>>_{\al}}{V}\\
f(x)=\frac{\int dy < \rho(y)>_{\al} < \rho (y+x>>_{\al}}{V}.
\eea

A correct description of the low temperature phase must take into account both correlation 
functions.  The replica method does it quite nicely: $g$ is the correlation function inside one 
replica and $f$ is the correlation function among two different replicas.
\bea
g(x)=\frac{\int dy < \rho_{a}(y) \rho _{a} (y+x>>}{V}\\
f(x)=\frac{\int dy < \rho_{a}(x) \rho _{b} (y+x>>}{V},
\eea
with $a \ne b$.

The problem is now to write closed equation for the two correlation functions $f$ and $g$.  
Obviously in the high temperature phase we must have that $f=0$ and the non-vanishing of $f$ is a 
signal of entering in the glassy phase.

The first attempt in this direction was only a partial success \cite{MEPA}.  A generalized 
hypernetted chain approximation was developed for the two functions \bd f \ed and $g$.  A non 
trivial solution was found at sufficient low temperature both for soft and hard spheres and the 
transition temperature to a glassy state was not very far from the numerically observed one.  
Unfortunately the value of the specific heat at low temperature is not the correct one (it strongly 
increases by decreasing the temperature).  Therefore the low temperature behaviour is not the 
correct one; this should be not a surprise because an esplicite computation show that the 
corrections to this hypernetted chain approximation diverge at low temperature.

These results show the feasibility of a replica computation for real glasses, however they point in 
the direction that one must use something different from a replicated version of the hypernetted 
chain approximation.  At the present moment it is not clear which approximation is the correct one, 
but I feel confident that a more reasonable one will be found in the near future.

\end{document}